\documentclass[sigconf]{acmart}

\copyrightyear{2020}
\acmYear{2020}
\setcopyright{licensedothergov}
\acmConference[MSR '20]{17th International Conference on Mining Software Repositories}{October 5--6, 2020}{Seoul, Republic of Korea}
\acmBooktitle{17th International Conference on Mining Software Repositories (MSR '20), October 5--6, 2020, Seoul, Republic of Korea}
\acmPrice{15.00}
\acmDOI{10.1145/3379597.3387510}
\acmISBN{978-1-4503-7517-7/20/05}

\usepackage[title,titletoc]{appendix}
\usepackage{algorithmic}
\usepackage{graphicx}
\usepackage{listings}
\usepackage{textcomp}
\usepackage{xcolor}
\usepackage{verbatim}
\usepackage[justification=centering]{caption}
\usepackage{subcaption}
\usepackage{siunitx}
\usepackage[utf8]{inputenc}

\usepackage{xspace}

\newcommand{\SWH}{Software Heritage\xspace}
\newcommand{\SWHGD}{\SWH Graph Dataset\xspace}

\newcommand{\gb}{\textsc{gb}\xspace}
\newcommand{\tb}{\textsc{tb}\xspace}
\newcommand{\sql}{\textsc{sql}\xspace}
\newcommand{\aws}{\textsc{aws}\xspace}

\newcommand{\figref}[1]{\figurename~\ref{#1}}

\definecolor{dkred}{RGB}{87,10,10}

\lstdefinelanguage[x]{sql}[]{sql}{morekeywords={WITH,DOUBLE,REGEXP_EXTRACT,FROM_UTF8,DAY_OF_WEEK,TO_BASE_64}}

\makeatletter
\newcommand\nocaption{\renewcommand\p@subfigure{}
    \renewcommand\thesubfigure{\thefigure\alph{subfigure}}
}
\makeatother

\title[The Software Heritage Graph Dataset]{The Software Heritage Graph Dataset:\\
  Large-scale Analysis of Public Software Development History}

\author{Antoine Pietri}
\email{antoine.pietri@inria.fr}
\orcid{0000-0003-4052-4469}
\affiliation{\institution{Inria}
  \city{Paris}
  \country{France}
}

\author{Diomidis Spinellis}
\email{dds@aueb.gr}
\orcid{0000-0003-4231-1897}
\affiliation{\institution{Athens University of Economics and Business}
  \city{Athens}
  \country{Greece}
}

\author{Stefano Zacchiroli}
\email{zack@irif.fr}
\orcid{0000-0002-4576-136X}
\affiliation{\institution{Université de Paris and Inria}
  \city{Paris}
  \country{France}
}

\begin{abstract}
  Software Heritage is the largest existing public archive of software source
  code and accompanying development history.  It spans more than five billion
  unique source code files and one billion unique commits, coming from more
  than 80 million software projects. These software artifacts were retrieved
  from major collaborative development platforms (e.g., GitHub, GitLab) and
  package repositories (e.g., PyPI, Debian, NPM), and stored in a uniform
  representation linking together source code
  files, directories, commits, and
  full snapshots of version control systems (VCS) repositories as observed by
  Software Heritage during periodic crawls. This dataset is unique in terms of
  accessibility and scale, and allows to explore a number of research questions
  on the long tail of public software development, instead of solely focusing
  on ``most starred'' repositories as it often happens.
\end{abstract}

\begin{document}
\maketitle

\section{Introduction}
\label{sec:intro}

Analyses of software development history have historically focused on crawling
specific ``forges''~\cite{DBLP:conf/wikis/Squire17} such as GitHub or GitLab,
or language specific package managers~\cite{DBLP:conf/msr/KikasGDP17,
  DBLP:conf/msr/AbateCGFTZ15}, usually by retrieving a selection of popular
repositories (``most starred'') and analyzing them individually. This approach
has limitations in scope: (1) it works on \emph{subsets} of the complete corpus
of publicly available software, (2) it makes \emph{cross-repository} history
analysis hard, (3) it makes \emph{cross-VCS} history analysis hard by not being
VCS-agnostic.

The \SWH project~\cite{cacm-2018-software-heritage, ipres-2017-software-heritage} aims to collect, preserve and share all the
  publicly available software source code, together with the associated
development history as captured by modern VCSs~\cite{spinellis2005version}. In
2019, we presented the \emph{\SWHGD}, a graph representation of all
the source code artifacts archived by \SWH~\cite{msr-2019-swh}.
The graph is a fully deduplicated
Merkle \textsc{dag}~\cite{Merkle} that contains the source code files,
directories, commits and releases of all the repositories in the archive.

The dataset captures the state of the \SWH archive on September 25th 2018,
spanning a full mirror of Github and GitLab.com, the Debian distribution,
Gitorious, Google Code, and the PyPI repository. Quantitatively it corresponds
to 5 billion unique file contents and 1.1 billion unique commits, harvested
from more than 80 million software origins (see Section~\ref{sec:data-model}
for detailed figures).

We expect the dataset to significantly expand the scope of software analysis by
lifting the restrictions outlined above: (1) it provides the best approximation
of the entire corpus of publicly available software, (2) it blends together
related development histories in a single data model, and (3) it abstracts over
VCS and package differences, offering a canonical representation (see
Section~\ref{sec:data-model}) of source code artifacts.

\section{Research questions and challenges}
\label{sec:rq}

The dataset allows to tackle currently under-explored research questions, and
presents interesting engineering challenges. There are different categories of
research questions suited for this dataset.

\begin{itemize}
\item \textbf{Coverage}: Can known software mining experiments be replicated
  when taking the distribution tail into account? Is language detection
  possible on an unbounded number of languages, each file having potentially
  multiple extensions? Can generic tokenizers and code embedding
  analyzers~\cite{Alon:2019} be built without knowing their
  language a priori?

\item \textbf{Graph structure}: How tightly coupled are the different layers of
  the graph? What is the deduplication efficiency across different
  programming languages? When do contents or directories tend to be reused?

\item \textbf{Cross-repository analysis}: How can forking and duplication
  patterns inform us on software health and risks?
  How can community forks be distinguished from personal-use
  forks? What are good predictors of the success of a community fork?

\item \textbf{Cross-origin analysis}: Is software evolution consistent across
  different VCS? Are there VCS-specific development patterns? How
  does a migration from a VCS to another affect development patterns? Is there
  a relationship between development cycles and package manager releases?

\end{itemize}

The scale of the dataset makes tackling some questions also an engineering
challenge: the sheer volume of data calls for distributed
computation, while analyzing a graph of this size requires state of the
art graph algorithms, being on the same order of magnitude as
WebGraph~\cite{BoVWFI,BRSLLP} in terms of edge and node count.

\appendix

\begin{appendices}

\section{Figures}
\label{sec:stats}

The dataset contains more than 11B software artifacts, as shown in
\figref{fig:figures}.

\begin{figure}[h]
  \begin{tabular}{l r}
    \hline\textbf{Table} & \textbf{\# of objects} \\ \hline
    origin    & \num{85143957}   \\
    snapshot  & \num{57144153}   \\
    revision  & \num{1125083793} \\
    directory & \num{4422303776} \\
    content   & \num{5082263206} \\ \hline
  \end{tabular}
  \caption{Number of artifacts in the dataset}
  \label{fig:figures}
\end{figure}

\section{Availability}
\label{sec:availability}

The dataset is available for download from Zenodo~\cite{msr-2019-swh} in two
different formats ($\approx$1~\tb each):

\begin{itemize}
\item a Postgres~\cite{stonebraker1991postgres} \textbf{database dump} in
  \textsc{csv} format (for the data) and \textsc{ddl} queries (for recreating
  the DB schema), suitable for local processing on a single server;

\item a set of \textbf{Apache Parquet files}~\cite{website-apache-parquet}
  suitable for loading into columnar storage and scale-out processing
  solutions, e.g., Apache Spark~\cite{zaharia2016apache}.
\end{itemize}

In addition, the dataset is ready to use on two different distributed cloud
platforms for live usage:

\begin{itemize}
  \item on \textbf{Amazon Athena},
    \cite{website-amazon-athena} which uses PrestoDB to distribute SQL queries.
  \item on \textbf{Azure Databricks};
    \cite{website-azure-databricks} which runs Apache Spark and can be queried
    in Python, Scala or Spark SQL.
\end{itemize}

\section{Data Model}
\label{sec:data-model}

\begin{figure}[h]
  \centering
  \includegraphics[width=0.8\linewidth]{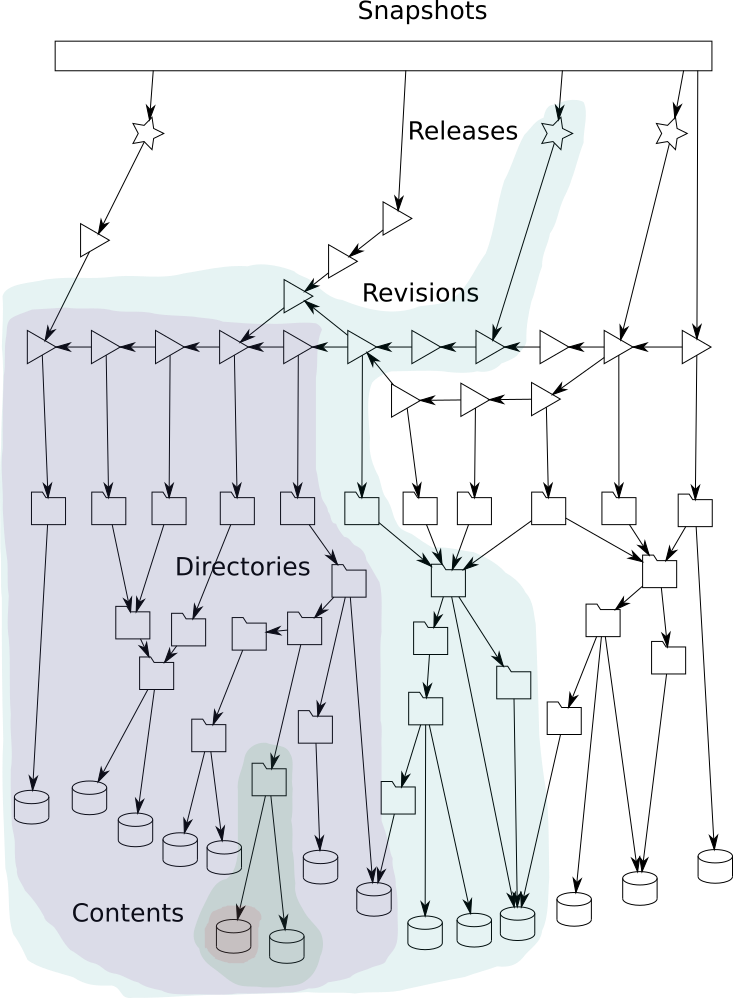}
  \caption{Data model: a uniform Merkle DAG containing source code artifacts
    and their development history}
  \label{fig:data-model}
\end{figure}

The \SWHGD exploits the fact that source code artifacts are massively
duplicated across hosts and projects~\cite{ipres-2017-software-heritage} to
enable tracking of software artifacts across projects, and reduce the storage
size of the graph.  This is achieved by storing the graph as a Merkle
directed acyclic graph (DAG)~\cite{Merkle}. By using persistent,
cryptographically-strong hashes as node identifiers~\cite{ipres-2018-doi},
the graph is deduplicated by sharing all identical nodes.

As shown in \figref{fig:data-model}, the \SWH DAG is organized in five logical
layers, which we describe from bottom to top.

\textbf{Contents} (or ``blobs'') form the graph's leaves,
and contain the raw content
of source code files, not including their filenames (which are
context-dependent and stored only as part of directory entries).
The dataset contains cryptographic checksums for all contents though,
that can be used to retrieve the actual files from any \SWH mirror using a Web
\textsc{api}\footnote{\url{https://archive.softwareheritage.org/api/}} and
cross-reference files encountered in the wild, including other datasets.

\textbf{Directories} are lists of named directory entries.
Each entry can
point to content objects (``file entries''), revisions (``revision entries''),
or other directories (``directory entries'').

\textbf{Revisions} (or ``commits'') are point-in-time captures of the entire
source tree of a development project. Each revision points to the root
directory of the project source tree, and a list of its parent revisions.

\textbf{Releases} (or ``tags'') are revisions that have been marked as noteworthy
with a specific, usually mnemonic, name (e.g., a version number). Each release
points to a revision and might include additional descriptive metadata.

\textbf{Snapshots} are point-in-time captures of the full state of a project
development repository. As revisions capture the state of a single development
line (or ``branch''), snapshots capture the state of \emph{all} branches in a
repository and allow to deduplicate unmodified forks across the archive.

Deduplication happens implicitly, automatically tracking byte-identical clones.
If a file or a directory is copied to another project, both projects will point
to the same node in the graph. Similarly for revisions, if a project is forked
on a different hosting platform, the past development history will be
deduplicated as the same nodes in the graph.  Likewise for snapshots, each
``fork'' that creates an identical copy of a repository on a code host, will
point to the same snapshot node.  By walking the graph bottom-up, it is hence
possible to find all \emph{occurrences} of a source code artifact in the
archive (e.g., all projects that have ever shipped a specific file content).

\begin{figure*}[t]
  \centering
  \includegraphics[height=0.35\textheight]{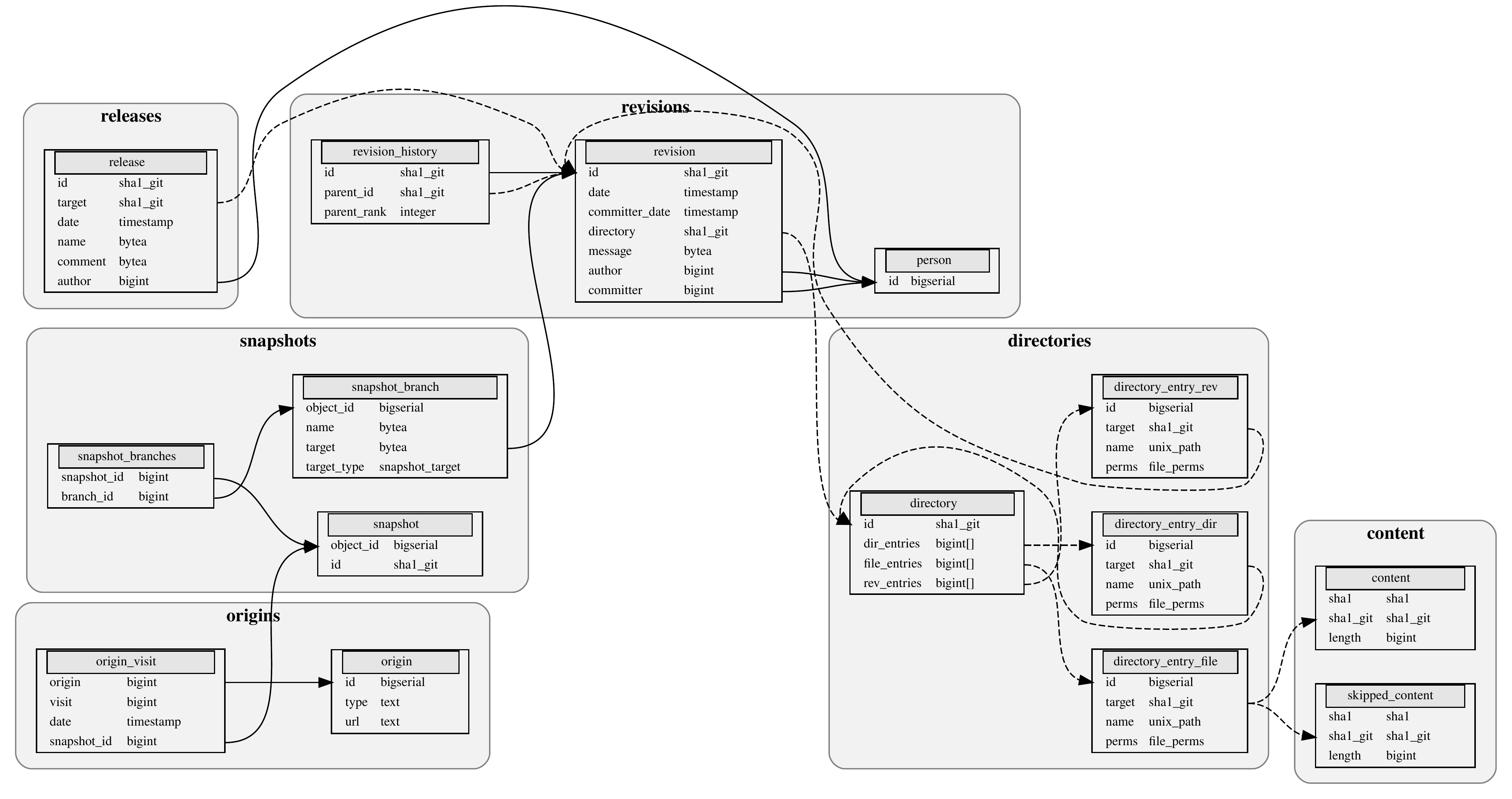}
  \caption{Simplified schema of the \SWHGD and the number of artifacts in it}\label{fig:db-schema}\end{figure*}

The Merkle \textsc{dag} is encoded in the dataset as a set of relational tables. In
addition to the nodes and edges of the graph, the dataset also contains
crawling information, as a set of triples capturing where (an origin \textsc{url}) and
when (a timestamp) a given snapshot has been encountered. A simplified view of
the corresponding database schema is shown in \figref{fig:db-schema}; the full schema
is available as part of the dataset distribution.

\section{Sample SQL Queries}
\label{sec:usage}

To further illustrate the dataset's affordances and as
motivating examples regarding the research possibilities
it opens,
below are some sample \sql\ queries that can be executed
with the dataset on \aws\ Athena.

\begin{lstlisting}[firstline=4,float=h,
caption={Most frequent file name},
label=lst:popular-file]
-- Most popular file name (in terms of number of different versions)
--  (Run time: 3 minutes 40 seconds, Data scanned: 150.68 GB)
-- cnt 182,373,697
SELECT FROM_UTF8(name, '?') AS name,
  COUNT(DISTINCT target) AS cnt
FROM directory_entry_file
GROUP BY name
ORDER BY cnt DESC
LIMIT 1;
\end{lstlisting}

Listing~\ref{lst:popular-file} shows a simple query
that finds the most frequent file name across all the revisions.
The result, obtained by scanning
151\gb\ in $3'40''$, is \texttt{index.html}, which occurs in the dataset 182
million times.

\begin{lstlisting}[firstline=3,float=h,
caption={Most common commit operations},
label=lst:popular-commit-words]
-- Fist word of 1.1 billion commit messages ordered by decreasing frequency
-- (Run time: 29.79 seconds, Data scanned: 37.51 GB)
SELECT COUNT(*) AS c, word
FROM
  (SELECT LOWER(REGEXP_EXTRACT(FROM_UTF8(
          message), '^\w+')) AS word
  FROM revision )
WHERE word != ''
GROUP BY  word
ORDER BY  COUNT(*) DESC LIMIT 20;
\end{lstlisting}

As an example of a query useful in software evolution research,
consider the Listing~\ref{lst:popular-commit-words}.
It is based on the convention dictating that commit messages should
start with a summary expressed in the imperative mood~\cite[3.3.2.1]{Fre19}.
Based on that idea, the query uses a regular expression to extract the first
word of each commit message and then tallies words by frequency.
By scanning 37~\gb\ in $30''$ it gives us that commits
concern the following common actions ordered by descending order of frequency:
\emph{add, fix, update, remove, merge, initial, create}.

\begin{lstlisting}[firstline=3,float=h,
caption={Ratio of commits performed during each year's weekends},
label=lst:weekend-work]
-- Percentage of weekend work over the years
--  (Run time: 7.55 seconds, Data scanned: 13.95 GB)
WITH revision_date AS
  (SELECT FROM_UNIXTIME(date / 1000000) AS date
  FROM revision)
SELECT yearly_rev.year AS year,
  CAST(yearly_weekend_rev.number AS DOUBLE)
  / yearly_rev.number * 100.0 AS weekend_pc
FROM
  (SELECT YEAR(date) AS year, COUNT(*) AS number
  FROM revision_date
  WHERE YEAR(date) BETWEEN 1971 AND 2018
  GROUP BY  YEAR(date) ) AS yearly_rev
JOIN
  (SELECT YEAR(date) AS year, COUNT(*) AS number
  FROM revision_date
  WHERE DAY_OF_WEEK(date) >= 6
      AND YEAR(date) BETWEEN 1971 AND 2018
  GROUP BY  YEAR(date) ) AS yearly_weekend_rev
  ON yearly_rev.year = yearly_weekend_rev.year
ORDER BY  year DESC;
\end{lstlisting}

\sql\ queries can also be used to express more complex tasks.
Consider the research hypothesis that weekend work on open source projects
is decreasing over the years as evermore
development work is done by companies rather than volunteers.
The corresponding data can be obtained by finding the ratio
between revisions committed on the weekends of each year and
the total number of that year's revisions (see Listing~\ref{lst:weekend-work}).
The results, obtained by scanning 14~\gb\ in $7''$ are inconclusive,
and point to the need for further analysis.

\begin{lstlisting}[firstline=4,float=h,
caption={Average number of a revision's parents},
label=lst:fork-size]
-- Average fork size of revisions
-- (Run time: 21.77 seconds, Data scanned: 23.33 GB)
-- Result is 1.0881200349675373
SELECT AVG(fork_size)
FROM (SELECT COUNT(*) AS fork_size
        FROM revision_history
        GROUP BY  parent_id);
\end{lstlisting}

The provided dataset forms a graph, which can be difficult query with \sql.
Therefore, questions associated with the graph's characteristics, such as
closeness, distance, and centrality, will require the use of other methods,
like Spark (see Section~\ref{sec:spark}).
Yet, interesting metrics can be readily obtained by limiting scans to specific
cases, such as merge commits.
As an example, Listing~\ref{lst:fork-size} calculates the average
number of parents of each revision ($1.088$, after scanning 23~\gb\ in $22''$)
by grouping revisions by their parent identifier.
Such queries can be used to examine in depth the characteristics
of merge operations.

\begin{lstlisting}[firstline=4,float,
caption={Average size of the most popular file types},
label=lst:file-type-size]
-- Average size of the most popular file types
-- calculated on a 1% sample from 5 billion files
-- (Run time: 1 minute 20 seconds, Data scanned: 317.16 GB)
SELECT suffix,
  ROUND(COUNT(*) * 100 / 1e6) AS Million_files,
  ROUND(AVG(length) / 1024) AS Average_k_length
FROM
  (SELECT length, suffix
  FROM
    -- File length in joinable form
    (SELECT TO_BASE64(sha1_git) AS sha1_git64, length
    FROM content ) AS content_length
  JOIN
  -- Sample of files with popular suffixes
  (SELECT target64, file_suffix_sample.suffix AS suffix
  FROM
    -- Popular suffixes
    (SELECT suffix FROM (
      SELECT REGEXP_EXTRACT(FROM_UTF8(name),
       '\.[^.]+$') AS suffix
    FROM directory_entry_file) AS file_suffix
    GROUP BY  suffix
    ORDER BY  COUNT(*) DESC LIMIT 20 ) AS pop_suffix
  JOIN
    -- Sample of files and suffixes
    (SELECT TO_BASE64(target) AS target64,
      REGEXP_EXTRACT(FROM_UTF8(name),
	'\.[^.]+$') AS suffix
    FROM directory_entry_file TABLESAMPLE BERNOULLI(1))
    AS file_suffix_sample
  ON file_suffix_sample.suffix = pop_suffix.suffix)
  AS pop_suffix_sample
  ON pop_suffix_sample.target64 = content_length.sha1_git64)
GROUP BY  suffix
ORDER BY  AVG(length) DESC;
\end{lstlisting}

Although the performance of Athena can be impressive,
there are cases where the available memory resources will be exhausted
causing an expensive query to fail.
This typically happens when joining two equally large tables consisting
of hundreds of millions of records.
This restriction can be overcome by sampling the corresponding tables.
Listing~\ref{lst:file-type-size} demonstrates such a case.
The objective here is to determine the modularity at the level of files among
diverse programming languages, by examining the size of popular file types.
The query joins two 5~billion row tables:
the file names and the content metadata.
To reduce the number of joined rows a 1\% sample of the rows is processed,
thus scanning 317~\gb\ in $1'20''$.
The order of the resulting language files
(JavaScript$>$C$>$C++$>$Python$>$\textsc{php}$>$C\#$>$ Ruby)
hints that,
with the exception of JavaScript,
languages offering more abstraction facilities are associated with
smaller source code files.

\section{Spark usage}
\label{sec:spark}

For a more fine-grained control of the computing resources, it is also possible
to use the dataset on Spark, through a local install or using the public
dataset on Azure Databricks.

Once the tables are loaded in Spark, the query in
Listing~\ref{lst:directory-outdegree} can be used to
generate an outdegree distribution of the directories.

\begin{lstlisting}[float=h,
caption={Outdegree distribution of directories},
label=lst:directory-outdegree]
%sql
select degree, count(*) from (
  select source, count(*) as degree from (
    select hex(source) as source,
           hex(target) as dest from (
      select id as source,
             explode(dir_entries) as dir_entry
      from directory)
    inner join directory_entry_file
      on directory_entry_file.id = dir_entry
  )
  group by source
)
group by degree
order by degree
\end{lstlisting}

To analyze the graph structure, the  GraphFrames
library~\cite{dave2016graphframes} can also be used to perform common
operations on the graph. Listing~\ref{lst:cc} demonstrates how one can load the
edges and nodes of the revision tables as a GraphFrame object, then compute the
distribution of the connected component sizes in this graph.

\begin{lstlisting}[float=h,
caption={Connected components of the revision graph},
label=lst:cc,language=python]
from graphframes import GraphFrame

revision_nodes = spark.sql("SELECT id FROM revision")
revision_edges = spark.sql("SELECT id as src, parent_id as dst "
                           "FROM revision_history")
revision_graph = GraphFrame(revision_nodes, revision_edges)

revision_cc = revision_graph.connectedComponents()
distribution = (revision_cc.groupby(['component']).count()
                .withColumnRenamed('count', 'component_size')
                .groupby(['component_size']).count())
display(distribution)
\end{lstlisting}

By allowing users to choose the amount of resources in the cluster, Spark lifts
the constraints imposed by limits in Athena, such as timeouts and limited
scale-out factor. This is important for computationally intensive
experiments or very large {\em join} operations,
which can only be achieved through sampling in Athena.

Spark is also more flexible in terms of the computations it can perform, thanks
to User-Defined Functions~\cite{website-spark-udf} that can be used to specify
arbitrary operations to be performed on the rows, which isn't possible with
Athena.

\section{Data sample}
\label{sec:sample}

A sample of the data as well as instructions to run live queries on the dataset
using Amazon Athena can be found at:
\url{https://annex.softwareheritage.org/public/dataset/graph/2018-09-25/}

\end{appendices}

\clearpage

\end{document}